\documentclass[11pt]{article}

\usepackage[latin1]{inputenc}
\usepackage[T1]{fontenc}
\usepackage[english]{babel}
\usepackage{amsfonts}
\usepackage{amssymb}
\usepackage{amsmath}
\usepackage{amsthm}
\usepackage{graphicx}

\newtheorem*{proposizione}{Proposition A}

\newtheorem*{teoremap}{Theorem A}

\def\be{\begin{equation}}
\def\ee{\end{equation}}
\def\bea{\begin{eqnarray}}
\def\eea{\end{eqnarray}}
\def\nin{\noindent}

\newcommand{\E}{\mathbb{E}}

\newtheorem{remark}{Remark}
\newtheorem{theorem}{Theorem}

\def\b{\beta}

\title{How glassy are neural networks?}

\author{Adriano Barra\footnote{Dipartimento di Fisica, Sapienza Universit\`a di Roma and GNFM, Sezione di Roma.},
Giuseppe Genovese\footnote{Dipartimento di  Matematica,
    Sapienza Universit\`a di Roma.}, \ Francesco Guerra\footnote{Dipartimento di Fisica, Sapienza Universit\`a di Roma and INFN, Sezione di Roma.}, Daniele Tantari\footnote{Dipartimento di Matematica, Sapienza Universit\`a di Roma.}}

\date{\today}

\begin{document}
\maketitle

\begin{abstract}
In this paper we continue our investigation on the high storage regime of a neural network with Gaussian patterns. Through an exact mapping between its partition function and one of a bipartite spin glass (whose parties consist of  Ising and Gaussian spins respectively), we give a complete control of the whole annealed region.
The strategy explored is based on an interpolation between the bipartite system and two independent spin glasses built respectively by dichotomic and Gaussian spins:
Critical line, behavior of the principal thermodynamic observables and their fluctuations as well as overlap fluctuations are obtained and discussed.
Then, we move further, extending such an equivalence beyond the critical line, to explore the broken ergodicity phase under the assumption of replica symmetry and we show that the quenched free energy of this (analogical) Hopfield model can  be described as a linear combination of the two quenched spin-glass free energies even in the replica symmetric framework.
\end{abstract}

\section*{Introduction}

Neural networks, thought of as the \textit{harmonic oscillators} of artificial intelligence, are nowadays being used in a huge number of different fields of science, ranging from practical application in data mining \cite{data1,data2} to theoretical speculation in systems biology \cite{JTB,enzo}, crossing fields as disparate as computer science \cite{montanari}, quantitative sociology \cite{epl} or economics \cite{ton}.
\newline
As a consequence, as applications develop, the need for mathematical methods (bringing them under rigorous control) and a simple mathematical framework (acting as a benchmark for future speculation) increases and motivates the present paper.
\newline
Moreover, although the Hopfield model has been extensively studied since it was introduced in \cite{hopfield}, both from a physical \cite{amit,ags2,isaac,peter, SVM} and a more mathematical \cite{tirozzi1,bovier2,bovier3,bovier4,tirozzi2,tirozzi3,talahopfield1,talahopfield2} point of view, from the rigorous perspective many points about its properties remain unsolved, which also prompts further efforts in developing new mathematical techniques and different physical perspectives.
%
\newline
In the past, we gave an extensive treatment of an analogical neural network \cite{NN1}\cite{NN2}, namely a mean-field structure with $N$ dichotomic neurons (spins) interconnected through Hebbian couplings \cite{amit,peter} whose $p$ patterns are stored according  to a standard Gaussian $\mathcal{N}[0,1]$: In  \cite{NN1} we studied its thermodynamical properties paying attention to the annealed approximation (but we were unable, at that time, to gain a complete control of the whole annealed region), while in \cite{NN2} we investigated the properties of the replica symmetric approximation.
\newline
Within our approach, the equilibrium statistical mechanics of the neural network is shown to be equivalent to the one of  a bipartite spin glass whose parts consist of the original $N$ neurons (belonging to the first party, hence made of by dichotomic variables) and the other hand $p$ Gaussians that give rise to the second part (hence consisting of continuous variables): As the theory of the mean-field Ising spin glass (namely the Sherrington Kirkpatrick model \cite{MPV}) has been intensively developed in the past decades (see for instance \cite{ALR}\cite{barra1}\cite{broken}\cite{limterm}),  while the same did not happen for the Gaussian counterpart, we investigated in detail the structure of the latter too, deepening the understanding of its properties in \cite{gauss1}.
\newline
Furthermore, to complete a streamlined description of the state of the art on this theme, we stress that results on the analogical Hopfield model, stemming from a mathematical perspective far from our connection with bipartite spin glasses, have also been obtained in \cite{bovier2,bovier3,bovier4}.
\newline
Turning to the applied side, despite the fact that in neural networks (in their original artificial intelligence framework) the interest in continuous patterns is reduced or moved to rotators (e.g. Kuramoto oscillators \cite{kuramoto}), as digital processing by Ising spins works as a better approximation for the standard \textit{integrate and fire} models of neurons \cite{stein}, in several other fields of science (as, for instance, in chemical kinetics \cite{enzo,aldo} or theoretical immunology \cite{JTB}) continuous values of patterns can instead be preferred (\cite{isaac}\cite{noest}) and a rigorous mathematical control of completely continuous models (namely with both continuous patterns and neurons) belongs to our strategy of research.
\newline
For the moment, we limit ourselves in presenting a clear scenario for the hybrid model made of by continuous patterns and dichotomic variables, namely the analogical neural network: In Section One we introduce the model and all the statistical-mechanics-related concepts. Then in Section Two we expose our new strategy of interpolation which allows a complete control both of the ergodic region (confirming the annealed approximation, which is investigated in great detail), and of the replica symmetric scenario, which is then deepened in Section Three.
\newline
The last section contains our conclusions.
\newline
Furthermore, an appendix is added: there the fluctuation theory of the order parameters of the model is discussed, and it is shown that the critical line found in this work characterizes a second order phase transition.

\section{The model, basic definitions and properties}

\subsection{The analogical Hopfield model}

We introduce a large network of $N$ two-state neurons  $ (1,..,N)\ni i \to \sigma_i = \pm1$, which are thought of as quiescent when their value is $-1$ or spiking when their value is $+1$. They interact
throughout a synaptic matrix $J_{ij}$ defined according to the Hebb
rule for learning \cite{hebb,hopfield}
\be J_{ij} = \sum_{\mu=1}^p \xi_i^{\mu}\xi_j^{\mu}. \ee
Each random variable $\xi^{\mu}=\{\xi_1^{\mu},..,\xi_N^{\mu}\}$
represents a learned pattern: While in the standard literature these patterns are usually chosen at random
independently with values $\pm1$ taken with equal probability $1/2$, 
we chose them as taking real values with a unit
Gaussian probability distribution, \textit{i.e.} \be d\mu(\xi_i^{\mu}) =
\frac{1}{\sqrt{2\pi}}e^{-(\xi_i^{\mu})^2/2}. \ee
The analysis of the network assumes that the
system has already stored $p$ patterns (no learning is
investigated here), and we will be interested in the case in which this
number asymptotically increases linearly with respect to the system size
(high storage level), so that $p/N\to\alpha$ as $N\to\infty$, where $\alpha\ge0$ is a parameter of the theory denoting the storage level.
\newline
The Hamiltonian of the model has a mean-field structure and involves interactions between any pair of sites according to the definition \be\label{ciao} H_N(\sigma;\xi) = - \frac{1}{N}\sum_{\mu=1}^p
\sum_{i<j}^N\xi_i^{\mu}\xi_j^{\mu}\sigma_i\sigma_j. \ee
\subsection{Morphism in the bipartite model}

By splitting the summations $\sum_{i<j}^N =
\frac{1}{2}\sum_{ij}^N - \frac12 \sum_i^N \delta_{ij}$ in the Hamiltonian (\ref{ciao}), we can introduce and write  the partition function $Z_{N,p}(\beta;\xi)$ in the following form
\begin{eqnarray}\label{due}
Z_{N,p}(\beta;\xi) &=& \sum_{\sigma}
\exp{\Big(\frac{\beta}{2N}\sum_{\mu=1}^p\sum_{ij}^N
\xi_i^{\mu}\xi_j^{\mu}\sigma_i\sigma_j -
\frac{\beta}{2N}\sum_{\mu=1}^p\sum_{i}^N (\xi_i^{{\mu}})^2 \Big)} \\ &=&
\tilde{Z}_{N,p}(\beta;\xi)e^{\frac{-\beta}{2N}\sum_{\mu=1}^p\sum_{i=1}^N
(\xi_i^{{\mu}})^2} \nonumber \end{eqnarray}
where $\beta\ge0$ is the inverse temperature, and denotes here the level of noise in the network. We have defined
\be\label{Ztilde} \tilde{Z}_{N,p}(\beta;\xi)=
\sum_{\sigma}\exp(\frac{\beta}{2N}\sum_{\mu=1}^p\sum_{ij}^N
\xi_i^{\mu}\xi_j^{\mu}\sigma_i\sigma_j ). \ee Notice that the last
term at the r.h.s. of eq. (\ref{due}) does not depend on the
particular state of the network, hence the control of the last term can be easily obtained \cite{NN1} and simply adds a factor $\alpha\beta/2$ to the free energy.
\newline
Consequently we focus just on $\tilde{Z}(\beta;\xi)$. Let us apply
the Hubbard-Stratonovich lemma \cite{ellis} to linearize with
respect to the bilinear quenched memories carried by the
$\xi_i^{\mu}\xi_j^{\mu}$.
\newline
We can write
\be\label{enne} \tilde{Z}_{N,p}(\beta;\xi)
 = \sum_{\sigma}\int  (\prod_{\mu=1}^p \frac{dz_{\mu}\exp(-z^2_{\mu}/2)}{\sqrt{2\pi}}) \exp(\sqrt{\beta/N}\sum_{i,\mu}\xi_i^{\mu}\sigma_i z_{\mu}).
 \ee
For a generic function $F$ of the neurons, we define the
Boltzmann state $\omega_{\beta}(F)$ at a given level of noise
$\beta$ as the average \be \omega_{\beta}(F) = \omega(F)=
(Z_{N,p}(\beta;\xi))^{-1}\sum_{\sigma}F(\sigma)e^{-\beta
H_N(\sigma;\xi)} \ee and often we will drop the subscript $\beta$
for the sake of simplicity. The $s$-replicated Boltzmann state
is defined as the product state $\Omega = \omega^1\times \omega^2 \times ... \times
\omega^s$, in which all the single Boltzmann states are at the same noise level $\beta^{-1}$ and share an identical
sample of quenched memories $\xi$. For the sake of
clearness, given a function $F$ of the neurons of the $s$ replicas
and  using the symbol $a \in [1,..,s]$ to label
replicas, such an average can be written as
\be \Omega(F(\sigma^1,...,\sigma^s)) =
\frac{1}{Z_{N,p}^s}\sum_{\sigma^1}\sum_{\sigma^2}...\sum_{\sigma^s}
F(\sigma^1,...,\sigma^s)\exp(-\beta \sum_{a=1}^s
H_N(\sigma^{a},\xi)). \ee
The average over the quenched memories will be denoted by
$\mathbb{E}$ and for a generic function of these memories $F(\xi)$
 can be written as \be \mathbb{E}[F(\xi)] = \int
(\prod_{\mu=1}^p \prod_{i=1}^N \frac{d
\xi_i^{\mu}e^{-\frac{(\xi_i^{\mu})^2}{2}}}{\sqrt{2\pi}})F(\xi)=
\int F(\xi)d\mu(\xi), \ee with $ \mathbb{E}[\xi_i^{\mu}]=0$
and $ \mathbb{E}[(\xi_i^{\mu})^2]=1$.
\newline
Hereafter we will often denote the average over the gaussian spins as $d\mu(z)$. We use the symbol $\langle . \rangle$ to mean $\langle . \rangle =
\mathbb{E}\Omega(.)$.
\newline
We recall that in  the thermodynamic limit it is assumed
$$
\lim_{N \rightarrow \infty} \frac{p}{N}= \alpha,
$$
$\alpha$ being a given real number, which acts as free parameter of the theory.

\subsection{The thermodynamical observables}

The main quantities of interest are the intensive pressure,
defined as \be \lim_{N \to \infty} A_{N,p}(\beta,\xi)= -\beta \lim_{N \to \infty} f_{N,p}(\beta,\xi) =\lim_{N \to \infty}
\frac{1}{N}\ln Z_{N,p}(\beta;\xi), \ee  the quenched intensive pressure,
 defined as
\be \lim_{N \to \infty} A^*_{N,p}(\beta)= -\beta \lim_{N \to \infty} f^*_{N,p}(\beta) =
\lim_{N \to \infty} \frac{1}{N}\mathbb{E}\ln Z_{N,p}(\beta;\xi), \ee and the
annealed intensive pressure, defined as  \be
\lim_{N \to \infty} \bar{A}_{N,p}(\beta) = -\beta \lim_{N \to \infty} \bar{f}_{N,p}(\beta) = \lim_{N \to \infty}
\frac{1}{N}\ln \mathbb{E} Z_{N,p}(\beta;\xi). \ee According to thermodynamics, here
$f_{N,p}(\beta,\xi)= u_{N,p}(\beta,\xi)-\beta^{-1}s_{N,p}(\beta, \xi)$
is the free energy density, $u_{N,p}(\beta,\xi)$ is the internal
energy density and $s_{N,p}(\beta,\xi)$ is the intensive entropy (the star and
the bar  denote the quenched and the annealed evaluations as
well).
\newline
According to the exploited bipartite nature of the Hopfield model,  we introduce two other order parameters:  the first is the overlap between the replicated neurons, defined as \be q_{ab}= \frac1N
\sum_{i=1}^N \sigma_i^a \sigma_i^b \in [-1,+1], \ee and the second
the overlap between the replicated Gaussian  variables $z$,
defined as \be p_{ab} = \frac1p \sum_{\mu=1}^p z_{\mu}^{a}z_{\mu}^{b}
\in (-\infty, +\infty). \ee
These overlaps play a considerable role in the theory
as they can express thermodynamical quantities.

\section{A detailed description of the annealed region}

\subsection{The interpolation scheme for the annealing}

In this section we present the main idea of the work, used here to get a complete control of the high-temperature region: We interpolate between the neural network (described in terms of a bipartite spin glass) and a system consisting of two separate spin glasses, one dichotomic and one Gaussian. Note that, by the Jensen inequality, namely
$$
\mathbb{E} \ln Z_{N,p}(\beta) \leq \ln \mathbb{E} Z_{N,p}(\beta),
$$
we can write
\be\label{bound}
A^{*}_{N,p} \leq \frac1N\ln \mathbb{E} \sum_{\sigma}\int \prod_{\mu=1}^p d\mu(z_{\mu}) e^{\sqrt{\frac{\beta}{N}}\sum_{i \mu}\xi_i^{\mu}\sigma_i z_{\mu}} = \ln 2 - \frac{p}{2N}\log(1-\beta),
\ee
where we emphasize that the integral inside eq. (\ref{bound}) exists only for $\beta < 1$.
\newline
The $N\to \infty$ limit then offers immediately $\lim_{N \to \infty}A^{*}_{N,p}(\beta) \leq \ln 2 - \alpha \ln(1-\beta)/2$.
The next step is to use interpolation to prove the validity of the Jensen bound in the whole region defined by the line $\beta_c = 1/(1+\sqrt{\alpha})$, which defines the boundary of the validity of the annealed approximation, in complete agreement with the well known picture of Amit, Gutfreund and Sompolinsky \cite{amit}\cite{ags2}.
\newline
To understand which is the proper interpolating structure, let us note that the exponent of the Boltzmann factor yields a family of random variables indexed by the configurations $(\sigma,z)$. For a given realization of the noise, $H(\sigma,z|\xi)=\sqrt{\frac{\beta}{N}}\sum_{i \mu} \xi_{i,\mu}\sigma_i z_{\mu}$ is a randomly centered variable with variance
$$
\mathbb{E}\large(H(\sigma,z|\xi)H(\sigma',z'|\xi)\large) = \frac{\beta}{N}\sum_{i \mu}\sigma_i \sigma_i^{\prime} z_{\mu}z_{\mu}^{\prime} = \beta p q_{\sigma \sigma^{\prime}} p_{z z^{\prime}}.
$$
The presence of the product $q_{\sigma \sigma^{\prime}} p_{z z^{\prime}}$ in the variance suggests the correct interpolating structure among this bipartite network and two other independent spin glasses, namely a Sherrington-Kirkpatrick  model with variance $q_{\sigma \sigma^{\prime}}^2$ and another spin glass model with Gaussian spin and variance $p_{z z^{\prime}}^2$. It is in fact clear that a proper interpolating structure can be held by
\begin{eqnarray}\label{piedi}
\varphi_N(t) &=& \frac1N \mathbb{E} \ln \sum_{\sigma} \int \prod_{\mu=1}^p d\mu(z_{\mu})
\exp{\Large( \sqrt{t}\sqrt{\frac{\beta}{N}}\sum_{i\mu}\xi_i^{\mu}\sigma_i z_{\mu}\Large)} \\ \nonumber
&\cdot& \exp{\Large( \sqrt{1-t}( \beta_1\sqrt{\frac{N}{2}} K(\sigma) + \beta_2 \sqrt{\frac{p}{2}} \bar{K}(z)) \Large)} \\ \nonumber
&\cdot& \exp{\Large( (1-t)( \frac{p \beta}{2}p_{z z} - \frac{p \beta^2_2}{4}p^2_{z z}) \Large)},
\end{eqnarray}
where we have set
$$
K(\sigma)=\frac{1}{N}\sum_{ij}J_{ij}\sigma_i\sigma_j
$$
and
$$
\bar K(z)=\frac{1}{p}\sum_{ij}\bar J_{ij}z_iz_j
$$
and the average $\E$ is taken with respect to all the i.i.d. normal random variables $\xi_{ij},J_{ij},\bar J_{ij}$. The interpolation is performed such that for $t=1$ the interpolating structure $\varphi(t=1)$ returns the free energy of the bipartite model, namely of the neural network, while for $t=0$ it coincides with a factorization in an SK spin glass and a (suitably regularized) Gaussian one \cite{gauss1}; $\beta_1, \beta_2$, which will be then fixed as  opportune noise levels, for the moment are simply free parameters.
\newline
As in \cite{NN2}\cite{Gsum}, the plan is now to evaluate the flow under a changing $t$ of the interpolating structure in order to get a positive defined sum rule by tuning opportunely $\beta_1,\beta_2$; hence, if we generalize the states as $\langle . \rangle_t = \mathbb{E}\Omega_t$, where the subscript $t$ accounts for the extended interpolating structure defined in (\ref{piedi}) we can write
\begin{eqnarray}
\frac{d \varphi_N(t)}{dt} &=& \frac1N \frac12 \beta p \Big( \langle p_{z z} \rangle_t - \langle q_{\sigma \sigma^{\prime}} p_{z z^{\prime}} \rangle_t \Big) - \frac{1}{4}\beta_1^2\Big( 1 - \langle q_{\sigma \sigma^{\prime}}^2 \rangle_t\Big) + \\
&-& \frac{p}{N}\frac{1}{4} \beta_2^2 \Big( \langle p_{z z}^2 \rangle_t - \langle p_{z z^{\prime}}^2 \rangle_t \Big)
+ \frac{p}{N}\frac{1}{4}\beta^2_2 \langle p_{zz}^2 \rangle_t - \frac{\beta}{2}\frac{p}{N}\langle p_{zz} \rangle_t,
\end{eqnarray}
then, calling $\alpha= p/N$ even at finite size $N$ (with a little language abuse), we can write
\begin{equation}
\frac{d \varphi_N(t)}{dt} = -\frac{\beta^2_1}{4}+\frac14 \langle \beta_1^2 q^2_{\sigma \sigma^{\prime}} + \alpha \beta_2^2 p_{z z^{\prime}}^2 - 2 \alpha \beta q_{\sigma \sigma^{\prime}} p_{z z^{\prime}} \rangle_t.
\end{equation}
If we now impose on $\beta_1, \beta_2$ the constraint $\beta_1 \beta_2 = \sqrt{\alpha}\beta$ we get a perfect square in the brackets of the  flow under a changing $t$, and calling $S_t(\alpha,\beta)= \langle (\beta_1 q_{\sigma \sigma^{\prime}} - \sqrt{\alpha}\beta_2 p_{z z^{\prime}})^2 \rangle_t$ the source term, we can write
\be\label{sumrule}
\frac{d \varphi_N}{dt} \geq - \frac14 \beta_1^2 + S_t(\alpha,\beta).
\ee
We can then integrate back between $[0,1]$ to get the following inequality
\begin{eqnarray}\label{t_annealed}
\varphi_N(1)&=&\frac1N \mathbb{E}\ln\sum_{\sigma}\int \prod_{\mu}^p d\mu(z_{\mu})e^{\sqrt{\frac{\beta}{N}}\sum_{i \mu}\xi_i^{\mu} \sigma_i z_{\mu}} \nonumber\\
&\geq& \frac1N \mathbb{E}\ln \sum_{\sigma} e^{\beta_1 \sqrt{\frac{N}{2}} K(\sigma)}\nonumber\\
&-& \frac{\beta^2_1}{4}+\frac{p}{N} \frac1p \mathbb{E}\ln \int \prod_{\mu}d\mu(z_{\mu}) e^{\beta_2 \sqrt{\frac{p}{2}}\bar{K}(z)} e^{-\frac{\beta^2_2 p}{4}p_{z z^{\prime}}}e^{\frac{p}{2}\beta p_{z z}},  \nonumber
\end{eqnarray}
under the constraint $\beta_1 \beta_2 = \sqrt{\alpha}\beta$.
\newline
Note that $K(\sigma)$ in the above expression defines the SK-model, while the last term defines the regularized Gaussian spin glass deeply investigated in \cite{gauss1}.
\newline
Now the advantages of this interpolation scheme become evident: As we have extremely satisfactory descriptions of the two independent models, namely the SK and the Gaussian spin glass, by these properties we can infer the behavior of the neural network (again thought of as the bipartite spin glass).
\newline
In particular, we know that  the free energies of each single part spin glass approach their annealed expression in the region where $\beta_1 \leq 1$ \cite{talabook} and $\beta + \beta_2 \leq 1$ \cite{gauss1}. Within this region, at the r.h.s. of eq. (\ref{t_annealed}) we get, in the thermodynamic limit, exactly $\ln 2 - (\alpha/2) \ln (1-\beta)$.
\newline
Furthermore, if $\alpha$ and $\beta$ respect the constraint $\beta(1+\sqrt{\alpha})\leq 1$, then finding $\beta_1, \beta_2$ such that the conditions $(A),(B),(C)$ hold, being
$$
\beta_1 \beta_2 = \sqrt{\alpha}\beta \ \  (A), \ \ \beta_1 \leq 1 \ \ (B), \ \ \beta + \beta_1 \leq  1\ \ (C),
$$
is certainly possible. In particular, using the SK critical behavior for the sake of simplicity, hence posing $\beta_1=1$, and setting $\beta_2= \sqrt{\alpha}\beta$, conditions $(A)$ and $(B)$ are automatically satisfied and, for the latter, being $\beta_2= \sqrt{\alpha}\beta$, we get
$$
\beta + \beta_2 \equiv \beta + \sqrt{\alpha}\beta = \beta(1+\sqrt{\alpha}) \leq 1,
$$
such that also condition $(C)$ is verified.
We can then state the following
\begin{theorem}
In the $\alpha,\beta$ plane there  exist a critical line, defined by
\be\label{critica}\beta_c(\alpha) = \frac{1}{1+\sqrt{\alpha}},\ee
such that for $\beta\le\beta_c(\alpha)$  the annealed approximation of the free energy holds
\be
\lim_{N \to \infty}\frac1N \mathbb{E}\ln \sum_{\sigma}\int \prod_{\mu} d\mu(z_{\mu}) e^{\Big( \sqrt{\frac{\beta}{N}}\sum_{i \mu}\xi_i^{\mu}\sigma_i z_{\mu}  \Big)} = \ln 2 - \frac{\alpha}{2} \ln(1-\beta).
\ee
\end{theorem}
\begin{remark}
We stress that the Borel-Cantelli lemma allows straightforwardly to determine the correct annealed regions for the SK model \cite{talabook} and, through a careful check of convergence of the integral defining the partition function, the same holds for the Gaussian case too \cite{gauss1}; however, the direct application of the Borel-Cantelli argument on the neural network gives a weaker result as shown for instance in \cite{NN1}. The interpolation scheme allows to exploit and transfer the results for the SK and Gaussian models to the neural network, and enlarges the area of validity of the annealed expression for the free energy to the whole expected region, obtained  e.g. via the replica method \cite{amit}.
\end{remark}

\subsection{The control of the annealed region}

As a consequence, we can now extend the previous results exposed in \cite{NN1} to the whole annealed region:
 Summarizing, we get the following
\begin{theorem}\label{intensive}
There exists  $\beta_{c}(\alpha)$, defined by eq. (\ref{critica}), such that for $\beta<\beta_{c}(\alpha)$ we have the following limits for the intensive free energy, internal energy and entropy, as $N\to\infty$ and $p/N\to\alpha>0$:
\begin{eqnarray}
-\beta \lim_{N\to\infty} f_{N,p}(\beta;\xi) &=& \lim_{N\to\infty}N^{-1}\ln Z_{N,p}(\beta;\xi)
\\ \nonumber &=&\ln 2 -(\alpha / 2) \ln (1-\beta)- (\alpha\beta / 2),\\
\lim_{N\to\infty} u_{N,p}(\beta;\xi) &=& -\lim_{N\to\infty}N^{-1}\partial_{\beta}\ln Z_{N,p}(\beta;\xi) \\ \nonumber &=& -\alpha\beta / (2(1-\beta)),\\
\lim_{N\to\infty} s_{N,p}(\beta;\xi) &=& \lim_{N\to\infty}N^{-1}(\ln Z_{N,p}(\beta;\xi)-\beta \partial_{\beta}\ln Z_{N,p}(\beta;\xi))\\ \nonumber &=& \ln 2 - (\alpha / 2)\ln (1-\beta) -
(\alpha\beta^{2})/(2(1-\beta)) - (\alpha\beta / 2),
\end{eqnarray}
$\xi$-almost surely.
The same limits hold for the quenched averages, so that in particular
$$\lim_{N\to\infty}N^{-1}\mathbb{E}\ln Z_{N,p}(\beta;\xi)=\ln 2 -\frac{\alpha}{2}\ln (1-\beta)-\frac{\alpha\beta}{2},$$
where, in all these formulas, the last term, namely $-\alpha\beta/2$, arises due to the diagonal contribution of the complete partition function (\ref{due}).
\end{theorem}
\begin{theorem}\label{fluctuations}
There exists  $\beta_{c}(\alpha)$, defined by eq. (\ref{critica}), such that  for $\beta<\beta_{c}(\alpha)$ we have the following convergence in distribution
\be \ln {\tilde Z}_{N,p}(\beta; \xi)-\ln \mathbb{E}{\tilde Z}_{N,p}(\beta; \xi)\rightarrow C(\beta) + \chi S(\beta)
\ee
where $\chi$ is a unit Gaussian in $\mathcal{N}[0,1]$ and
\begin{eqnarray}
C(\beta) &=& -\frac{1}{2}\ln\sqrt{1/(1-\sigma^2\beta^2\alpha)} \\
S(\beta) &=& \Big( \ln \sqrt{1/(1-\sigma^2\beta^2\alpha)}\Big)^{\frac{1}{2}},
\end{eqnarray}
with $\sigma=(1-\beta)^{-1}$.
\end{theorem}

\section{Extension to the replica symmetric solution}

Once the correct interpolating structure is understood, and spurred by the observation that the replica symmetric expression for the quenched free energy of the three models, namely the analogical neural network, the SK spin glass and the Gaussian one, are well known and investigated (for instance in \cite{Gsum}\cite{barra1}\cite{mech}\cite{NN1}\cite{gauss1}) we want to push further the equivalence among neural network and spin glasses, giving a complete picture also of the replica symmetric approximation.
\newline
To this task, let us recall that the replica symmetric approximation of the quenched free energy of the analogical neural network $A_{NN}^{RS}(\alpha,\beta)$ is given by the following expression \cite{NN1}
\begin{eqnarray}\nonumber
A_{NN}^{RS}(\alpha,\beta) &=& \ln 2 + \int  d\mu(z) \ln\cosh \large( z \sqrt{\alpha \beta \bar{p}} \large) + \frac{\alpha}{2}\ln\large( \frac{1}{1-\beta (1-\bar{q})} \large) + \\ &+& \frac{\alpha\beta}{2} \frac{\bar{q}}{1-\beta(1-\bar{q})} - \frac{\alpha \beta}{2}\bar{p}(1-\bar{q}),
\end{eqnarray}
where the order parameters denoted with a bar (to mean their RS approximation) are given by
\begin{eqnarray}\label{qNN}
\bar{q} &=& \int  d\mu(z)  \tanh^2\Big( z \sqrt{\alpha \beta \bar{p}} \Big), \label{Qbip}\\
\bar{p} &=& \beta \bar{q} / \Big( 1-\beta(1-\bar{q}) \Big)^2.\label{pNN}
\end{eqnarray}
Let us introduce further $\beta_1$ and $\beta_2$ as
\begin{eqnarray}
\beta_1 &=& \frac{\sqrt{\alpha} \beta}{1 - \beta(1- \bar{q})}, \label{b1}\\
\beta_2 &=& 1 - \beta(1-\bar{q}),\label{b2}
\end{eqnarray}
such that $\beta_1 \beta_2 = \sqrt{\alpha}\beta$.
\bigskip
We need also the RS approximation $A_{SK}^{RS}(\beta_1)$ of the quenched free energy of the SK model, at the noise level $\beta_1$, namely
\be
A_{SK}^{RS}(\beta_1) = \ln 2 + \int d \mu(z) \ln \cosh \large( \beta_1 \sqrt{\bar{q}_{SK}} z \large) + \frac14 \beta_1^2\large( 1 - \bar{q}_{SK} \large)^2,
\ee
where
\be\label{Qsk}
\bar{q}_{SK} = \int d\mu(z) \tanh^2\Big( \beta_1 z \sqrt{\bar{q}_{SK}} \Big).
\ee
By a direct comparison among the overlap expressions (\ref{Qbip}, \ref{Qsk}) we immediately conclude that we must have
$$
\beta_1^2 \bar{q}_{SK} = \alpha \beta \bar{p},
$$
which indeed holds as it can be verified easily, bearing in mind the expression (\ref{pNN}) and (\ref{b1}) for $\bar{p}$ and $\beta_1$.
\newline
As a last ingredient we need to introduce also the replica symmetric expression $A_{Gauss}^{RS}(\beta_2,\b)$ of the Gaussian spin glass at a  noise level $\beta_2$ as \cite{gauss1}
\be
A_{Gauss}^{RS}(\beta_2,\b) = \frac12 \ln \sigma + \frac12 \beta_2^2 \bar{p}_{G} \sigma^2 + \frac14 \beta_2^2 \bar{p}_{G}^2,
\ee
where
\begin{eqnarray}\label{Pgsg}
\bar{p}_{G} &=& (\beta_2 - (1-\beta))/\beta_2^2, \\
\sigma^2 &=& 1/(1- \beta + \beta^2 \bar{p}_{G}).
\end{eqnarray}
Note that the definition of the overlap between continuous variables encoded by eq. (\ref{pNN}) is in perfect agreement with the same overlap defined within the framework of eq.(\ref{Pgsg}), because, being $\beta_2 = 1-\beta(1-\bar{q})$, we can write
\be
\bar{p}_{Gauss} = \frac{\beta_2 - (1-\beta)}{\beta_2^2} = \frac{1-\beta(1-\bar{q})-(1-\beta)}{(1- \beta(1-\bar{q}))^2}
= \frac{\beta \bar{q}}{(1-\beta(1-\bar{q}))^2}.
\ee
As a consequence, through a direct verification by comparison (that we omit as it is long and straightforward), we can state the final theorem of the paper:
\begin{theorem}
Fixed, at noise level $\beta$, $\b_1$ and $\b_2$ as in (\ref{b1}) and (\ref{b2}), the replica symmetric approximation of the quenched free energy of the analogical neural network can be linearly decomposed in terms of the replica symmetric approximation of the Sherrington-Kirkpatrick quenched free energy, at noise level $\beta_1$, and the replica symmetric approximation of the quenched free energy of the Gaussian spin glass, at noise level $\beta_2$, such that
\be
A_{NN}^{RS}(\beta) = A_{SK}^{RS}(\beta_1) -\frac14 \beta_1^2 + \alpha A_{Gauss}(\beta_2,\b),
\ee
and the inequality (\ref{t_annealed}) becomes an identity for the RS behavior.
\end{theorem}
\begin{remark}
We stress that the above Theorem is in agreement with the sum rule (\ref{sumrule}) of Section 2 as, in the replica symmetric approximation, $q_{\sigma \sigma^{\prime}} = \bar{q}$ and $p_{z z^{\prime}}= \bar{p}$, hence
\be
\beta_1 \bar{q} - \sqrt{\alpha}\beta_2 \bar{p} = \frac{\sqrt{\alpha}\beta\bar{q}}{(1-\beta(1-\bar{q}))^2}
-\sqrt{\alpha}\Big( 1-\beta(1-\bar{q})\Big)\frac{\beta \bar{q}}{(1-\beta(1-\bar{q}))^2}= 0.
\ee
\end{remark}
\begin{remark}
Approaching the high-temperature region we have $\bar{q} \to 0$ and $\bar{p} \to 0$, and clearly $\beta \to 1/(1+\sqrt{\alpha})$. As a consequence we have
\begin{eqnarray}
\beta_2 &=& 1 - \beta(1-\bar{q}) \to 1 - 1/(1+\sqrt{\alpha}), \\
\beta_1 &=& \frac{\sqrt{\alpha}\beta}{1-\beta(1-\bar{q})} \to 1,
\end{eqnarray}
then $\beta + \beta_2 \to 1$, such that also the single-party counterparts approaches their critical points.
\end{remark}
Coherently, inside the annealed region we get $\bar{q} = 0$, then with the expressions for $\beta_1, \beta_2$ we can write $\beta_2 + \beta = 1$ that is the boundary of the annealed region for the Gaussian spin glass, while $\beta_1 = \sqrt{\alpha}\beta/(1-\beta)$ because $\beta \leq 1/(1+\sqrt{\alpha})$ we get $\beta_1 \leq 1$, namely the annealed region of the SK model.

\section{Conclusions and Outlook}

Neural networks are becoming the paradigm of a wide family of complex systems with \textit{cognitive capabilities} such as memory and learning both in the living world and outside.
\newline
As a consequence, a solid control of these networks is fundamental: In this paper we provided a clear analysis of the analogical neural network thought of as a bipartite spin glass, made of by two different type of spins: one ensemble of dichotomic variables, as in the celebrated Sherrington-Kirkpatrick model, and one ensemble made of by Gaussian distributed variables.
\newline
Exploiting this analogy, we developed a new interpolation scheme among the bipartite spin glass that mirrors the neural network and two independent glassy systems. Through this novel technique, we have then shown how to get a complete control of the annealed region of the neural network: The critical line has been obtained, together with an explicit behavior of all the main thermodynamical quantities: free energy, internal energy, entropy and overlaps (namely the order parameters of the theory).
\newline
One step forward we extended our interpolation scheme beyond the ergodic region, under the assumption of replica symmetry: We showed that the replica symmetric approximation of the quenched free energy of the analogical Hopfield model (at noise level $\beta$) can be expressed in terms of the replica symmetric expressions of the quenched free energies of the SK model (at noise level $\beta_1$) and of the Gaussian model (at noise level $\beta_2$), and we obtained the equations linking $\beta,\beta_1,\beta_2$ obtaining then a complete control also within this framework.
\newline
All that opens very interesting perspectives. The structure of the neural network as a linear combination of spin glasses is very rich: in fact we know that, as the SK model presents a very glassy full RSB structure \cite{broken}, in the Gaussian one this is absent, since the true solution is in fact RS even with no external field \cite{gauss1}. Thus one could aspect in our analogical neural network a competition of these two effects: rather a new feature in the complex systems scenario, that has to be deeply investigated.
\newline
Clearly we would deepen this topic, for example within a fully broken replica symmetry scenario on which we plan to report soon.
\newline
Furthermore, the analogical model shares many features with the original Hopfield model (which is even harder from a mathematical point of view) for which one could study in what measure this structure is preserved.
\newline
Future outlooks should cover also the completely analogical model in order to develop mathematical techniques beyond the standard ones required in artificial intelligence and closer to system biology.

\bigskip
\nin {\bf Acknowledgements\\}
\nin The strategy of this work is founded by the MIUR trough the FIRB grant RBFR08EKEV which is acknowledged, together with Sapienza Universit\`a di Roma for partial financial support. Partial support from INFN is also acknowledged.

\section*{Appendix.\\Fluctuation Theory for the Order Parameters}

We develop in this appendix a fluctuation theory of the order parameters to see that the ergodicity breaking is accomplished through a second order phase transition (i.e. the overlap fluctuations, properly rescaled over the volume, do diverge on the line $\beta_c(\alpha)$ hence defining a critical phenomenon).
\newline
To satisfy this task we  proceed as follows: at first we introduce a different interpolating structure with respect to the one discussed above (developed and discussed in \cite{NN2}) to bridge the neural network with two single party one-body models where spins are subjected to random fields in a way close to stochastic stability \cite{sollich} or cavity perspective \cite{guerra2}.
Then we evaluate the flow with respect to the interpolating parameter so to be able to calculate variations of
generic observable as overlap correlation functions.
\newline
Then we define the centered and rescaled overlaps and introduce
their correlation matrix. Each element of this matrix then is
evaluated at $t=0$ and then propagated thought $t=1$ via its
flow: This procedure encodes naturally for a system of
coupled linear differential equations that, once solved, gives the
expressions of the overlap fluctuations. The latter are found to
diverge on the critical line $\beta_c(\alpha)$ already outlined and this will close our inspection of the annealed regime.
\newline
Let us start the plan by introducing the next interpolating structure:
\newline
In a pure stochastic stability fashion \cite{NN2}, we need to
introduce also two classes of i.i.d. $\mathcal{N}[0,1]$ variables,
namely $N$ variables $\eta_i$ and $p$ variables
$\tilde{\eta}_{\mu}$, whose average is still encoded into the
$\mathbb{E}$ operator and by which we define the following
interpolating quenched pressure $\tilde{\varphi}_{N,p}(\beta,t)$
\begin{eqnarray}\label{interpolante} &&\tilde{\varphi}_{N,p}(\beta,t) =
\frac1N\mathbb{E}\log\sum_{\sigma}\int \prod_{\mu}^p
d\mu(z_{\mu})\exp(\sqrt{t}\sqrt{\frac{\beta}{N}}\sum_{i,\mu}\xi_i^{\mu}\sigma_i
z_{\mu}) \\ &\cdot& \nonumber \exp(a\sqrt{1-t} \sum_i \eta_i
\sigma_i)\exp(b\sqrt{1-t}\sum_{\mu}
\tilde{\eta}_{\mu}z_{\mu})\exp(c\frac{(1-t)}{2}\sum_{\mu} z_{\mu}^2),
\end{eqnarray}
where
$$
a = \sqrt{\alpha \beta \bar{p}}, \ \ b = \sqrt{\beta \bar{q}} \ \
c = \beta (1 - \bar{q}).
$$
We stress that $t\in[0,1]$ interpolates between $t=0$ where the
interpolating quenched pressure becomes made of by non-interacting
systems (a series of one-body problem) whose integration is
straightforward (as well as the evaluation of the overlap correlation functions it produces)
and the opposite limit, $t=1$, that recovers the
correct quenched free energy.
Then we can evaluate the flow with respect to the Boltzman factor encoded in the structure (\ref{interpolante})
as stated in the next
\begin{proposizione}
Given $O$ as a smooth function of $s$ replica overlaps
$(q_1,...,q_s)$ and $(p_1,...,p_s)$, the following streaming
equation holds:
\begin{eqnarray} \label{streaming} \frac{d}{dt}\langle O \rangle_t
&=& \beta \sqrt{\alpha} \Big( \sum_{a,b}^s \langle O \cdot
\xi_{a,b}\eta_{a,b} \rangle_t \\ \nonumber &-& s \sum_{a=1}^s
\langle O \cdot \xi_{a,s+1}\eta_{a,s+1} \rangle_t +
\frac{s(s+1)}{2}\langle O \cdot \xi_{s+1,s+2}\eta_{s+1,s+2}
\rangle_t \Big). \end{eqnarray}
\end{proposizione}
We skip the proof as it is long but simple and works by a direct
evaluation which is pretty standard in the disordered system literature
(see for example \cite{Gsum,barra1}).

\bigskip

The rescaled overlap $\xi_{12}$ and $\eta_{12}$ are defined
accordingly to
\begin{eqnarray}
\xi_{12}= \sqrt{N}\Big( q_{12} - \bar{q} \Big), \\
\eta_{12}= \sqrt{K}\Big( p_{12} - \bar{p} \Big).
\end{eqnarray}

In order to control the overlap fluctuations, namely $\langle
\xi_{12}^2 \rangle_{t=1}$, $\langle \xi_{12}\eta_{12}
\rangle_{t=1}$, $\langle \eta_{12}^2 \rangle_{t=1}$,  ..., noting that
the streaming equation pastes two replicas to the ones already
involved ($s=2$ so far), we need to study nine correlation
functions. It is  then useful to introduce them and link them to
capital letters so to simplify their visualization:

\begin{eqnarray}
\langle \xi_{12}^2 \rangle_t &=& A(t), \ \ \  \langle
\xi_{12}\xi_{13}
\rangle_t = B(t), \ \ \ \langle \xi_{12}\xi_{34}\rangle_t = C(t), \\
\langle \xi_{12}\eta_{12} \rangle_t &=& D(t), \ \ \ \langle
\xi_{12}\eta_{13} \rangle_t = E(t), \ \ \ \langle
\xi_{12}\eta_{34}\rangle_t = F(t), \\
\langle \eta_{12}\eta_{12} \rangle_t &=& G(t), \ \ \ \langle
\eta_{12}\eta_{13} \rangle_t = H(t), \ \ \ \langle
\eta_{12}\eta_{34}\rangle_t = I(t).
\end{eqnarray}
If we introduce the operator \textit{dot} as
$$
\dot{O}  = \frac{1}{\beta\sqrt{\alpha}}\frac{d O}{dt},
$$
which simplifies calculations and shifts the propagation of the
flow from $t=1$ to $t=\beta\sqrt{\alpha}$.
Assuming a Gaussian behavior, as in the strategy outlined in \cite{Gsum}, we can write the overall flow of the overlap correlation functions in the form of  the following differential system
\begin{eqnarray}\nonumber
\dot{A} &=& 2AD - 8BE + 6CF, \\ \nonumber \dot{B} &=& 2AE + 2BD -
4BE - 6BF - 6EC + 12CF, \\ \nonumber \dot{C} &=& 2AF + 2CD + 8BE -
16BF - 16CE + 20CF, \\ \nonumber \dot{D} &=& AG - 4BH + 3CI + D^2
-4E^2 + 3F^2, \\ \nonumber \dot{E} &=& AH+BG -2BH -3BI -3CH + 6CI
+ 2ED -2E^2 -6EF + 6F^2, \\ \nonumber\dot{F} &=& AI + CG + 4BH
-8BI -8 CH + 10 CI  + 2DF + 4E^2 -16EF + 10F^2, \\ \nonumber
\dot{G} &=& 2GD - 8HE + 6IF, \\ \nonumber \dot{H} &=& 2GE + 2HD -
4HE - 6HF -6IE + 12IF, \\ \nonumber \dot{I} &=& 2GF + 2DI + 8HE -
16 HF - 16IE + 20IF.
\end{eqnarray}
 Although it may appear complex, it is relatively easy to solve this system, once the initial conditions at $t=0$ are known (information then can be
obtained straightforwardly as at $t=0$ everything factorizes the theory being one-body). Our general analysis covers also the case where external fields are involved. We do not report here the full analysis, for the sake of brevity.
\newline
Here, as we are interested in finding where ergodicity becomes broken,  we start
propagating  $t \in 0 \to 1$ from the annealed region, where $\bar{q}=0$ and $\bar{p}=0$, which simplifies further the problem:
\newline
In fact, it is immediate to check that, for the only terms that we need to
consider, $A,D,G$ (the other being strictly zero on the whole
$t\in [0,1]$), the starting points are $A(0)=1, D(0)=0,
G(0)=(1-\beta)^{-2}$ and their evolution is ruled by
\begin{eqnarray}
\dot{A} &=& 2AD, \\ \dot{D} &=& AG + D^2, \\ \dot{G} &=& 2GD.
\end{eqnarray}
The solution of this differential system is long but straightforward then we skip the proof and directly state the next
\begin{teoremap}
In the ergodic region the behavior of the overlap fluctuations is
regular and described by  the following equations
\begin{eqnarray}
\langle \xi_{12}^2 \rangle &=& \frac{(1-\beta)^2}{(1-\beta)^2 -
\beta^2 \alpha}, \\
\langle \xi_{12}\eta_{12} \rangle &=& \frac{
\beta\sqrt{\alpha}}{(1-\beta)^2 - \beta^2 \alpha}, \\
\langle \eta_{12}^2 \rangle &=& \frac{1}{(1-\beta)^2 - \beta^2
\alpha},
\end{eqnarray}
diverging on the critical line $\beta_{c}(\alpha)$, defined by eq. (\ref{critica}),
hence defining a second order phase transition.
\end{teoremap}


\begin{thebibliography}{9}


\bibitem{kuramoto} J.A. Acebron, L. Bonilla, C. Perez-Vicente, Prez; F. Ritort, R. Spigler, {\em The Kuramoto model: a simple paradigm for synchronization phenomena}, Rev. Mod. Phys. \textbf{77}, 137, (2005).

\bibitem{JTB} E. Agliari, A. Barra, F. Guerra, F. Moauro, {\em A thermodynamical perspective of immune capabilities}, J. Theor. Biol. \textbf{287}, 48, 2011.

\bibitem{ALR} M. Aizenman, J. Lebowitz, D. Ruelle, {\em Some Rigorous Results on the Sherrington-Kirkpatrick Model
of Spin Glasses}, Commun. Math. Phys., \textbf{112} 3-20 (1987).

\bibitem{tirozzi1} S. Albeverio, B. Tirozzi, B. Zegarlinski {\em Rigorous results for the free energy in the
Hopfield model}, Comm. Math. Phys. \textbf{150}, 337 (1992).

\bibitem{amit} D.J. Amit, {\em Modeling brain function: The world of attractor neural
network}, \ Cambridge University Press, (1992).


\bibitem{ags2} D.J. Amit, H. Gutfreund, H. Sompolinsky {\em Storing infinite numbers of patterns in a spin glass model of neural networks}, Phys. Rev. Lett. \textbf{55}, 1530-1533,  (1985).


\bibitem{barra1} A. Barra,
\textit{Irreducible free energy expansion and overlap locking in
mean field spin glasses}, J. Stat. Phys. \textbf{123}, 601-614
(2006).


\bibitem{epl} A. Barra, P. Contucci, {\em  Toward a quantitative approach to migrants integration}, Europhys. Lett. \textbf{89}, 68001, (2010).

\bibitem{NN1} A. Barra, F. Guerra, {\em About the ergodic regime in the analogical Hopfield neural networks. Moments of the partition function}, J. Math. Phys. \textbf{49}, 125217 (2008).

\bibitem{NN2} A. Barra, G. Genovese, F. Guerra, {\em The replica symmetric behavior of the analogical neural network}, J. Stat. Phys. \textbf{140}, 784, (2010).

\bibitem{gauss1} A. Barra, G. Genovese, F. Guerra, D. Tantari, {\em A solvable mean field model of a Gaussian spin glass}, submitted, available at arxiv:1109.4069.


\bibitem{Bip} A.Barra, G.Genovese, F.Guerra, \textit{Equilibrium statistical mechanics of bipartite spin systems},
J. Phys. A: Math. and Theor. \textbf{44},  245002, (2011).

\bibitem{data1} J. P. Bigus, {Data Mining With Neural Networks: Solving Business Problems from Application Development to Decision Support}, McGraw-Hill (2006).


\bibitem{isaac} D. Boll\'e, T. M. Nieuwenhuizen, I. Perez-Castillo, T . Verbeiren, {\em A spherical Hopfield model},
J. Phys. A \textbf{36}, 10269, (2003).


\bibitem{bovier2} A. Bovier, A.C.D. van Enter and B. Niederhauser, {\em Stochastic symmetry-breaking in a Gaussian Hopfield-model}, J. Stat. Phys.  \textbf{95}, 181-213 (1999).

\bibitem{bovier3} A. Bovier, V. Gayrard {\em An almost sure central limit theorem for the Hopfield model}, Markov Proc. Rel. Fields  \textbf{3}, 151-173 (1997).

\bibitem{bovier4} A. Bovier {\em Self-averaging in a class of generalized Hopfield models}, J. Phys. A \textbf{27}, 7069-7077 (1994).

\bibitem{stein} A. N. Burkitt, {\em A review of the integrate-and-fire neuron model}, Biol Cybern \textbf{95}, 1, (2006).


\bibitem{ton} A.C.C. Coolen, {\em The mathematical theory of minority games: Statistical mechanics of interacting agents}, Oxford University Press,  (2005).

\bibitem{noest} A.C.C. Coolen, A.J. Noest, G.B. de Vries, {\em Modelling Chemical Modulation of Neural Processes},
Network \textbf{4}, 101, (1993).

\bibitem{peter} A.C.C. Coolen, R. Kuehn, P. Sollich, {\em Theory of Neural Information Processing
Systems}, Oxford University Press, 2005.

\bibitem{enzo} D. De Martino, M. Figliuzzi, A. De Martino, E. Marinari, {\em Computing fluxes and chemical potential distributions in biochemical networks: energy balance analysis of the human red blood cell}, submitted. Available at arxiv:1107.2330 (2012).

\bibitem{aldo} A. Di Biasio, E. Agliari, A. Barra, R. Burioni, {\em Cooperativity in chemical kinetics}, Theor. Chem. Acc. \textbf{131}, 1104, (2012).

\bibitem{SVM} R. Dietrich, M. Opper, and H. Sompolinsky. {\em Statistical mechanics of support vector networks},
Phys. rev. lett. \textbf{82}, 2975, (1999).

\bibitem{ellis} R.S. Ellis,
{\em Large deviations and statistical mechanics}, Springer, New
York, 1985.

\bibitem{mech} G. Genovese, A. Barra, {\em A mechanical approach to mean field spin models}, J. Math. Phys. \textbf{50}, 365234 (2009).

\bibitem{guerrasg} F. Guerra, {\em An introduction to mean field spin glass theory: methods and results},
In: \textit{Mathematical Statistical Physics}, A. Bovier et al. eds,
$243-271$, Elsevier, Oxford, Amsterdam, 2006.


\bibitem{broken} F. Guerra, {\em Broken Replica Symmetry Bounds in the
Mean Field Spin Glass Model}, Commun, Math. Phys. \textbf{233:1},
1-12 (2003).

\bibitem{guerra2} F. Guerra, {\em About the overlap distribution in mean field
spin glass models}, Int. Jou. Mod. Phys. B {\bf 10}, 1675-1684
(1996).

\bibitem{Gsum} F. Guerra, {\em Sum rules for the free energy in the mean
field spin glass model}, in {\em Mathematical Physics in
Mathematics and Physics: Quantum and Operator Algebraic Aspects},
Fields Institute Communications {\bf 30}, American Mathematical
Society (2001).

\bibitem{limterm} F. Guerra, F. L. Toninelli, {\em
The Thermodynamic Limit in Mean Field Spin Glass Models}, Commun.
Math. Phys. {\bf 230:1}, 71-79 (2002).



\bibitem{hebb} D.O. Hebb, {\em Organization of Behaviour}, Wiley, New York, 1949.




\bibitem{hopfield} J.J. Hopfield, {\em Neural networks and physical systems with emergent
collective computational abilities}, Proc. Ntl. Acad. Sci. USA
\textbf{79},  2554-2558 (1982).

\bibitem{montanari} M. M\'ezard, A. Montanari, {\em Information, Physics and Computation}, Oxford University press, 2009.

\bibitem{MPV} M. M\'ezard, G. Parisi and M. A. Virasoro, {\em Spin glass theory
and beyond}, World Scientific, Singapore, 1987.



\bibitem{tirozzi2} L. Pastur, M. Scherbina, B. Tirozzi, {\em The
replica symmetric solution of the Hopfield model without replica
trick} J. Stat. Phys. \textbf{74}, 1161-1183 (1994).

\bibitem{tirozzi3} L.Pastur, M. Scherbina, B. Tirozzi, {\em On the replica
 symmetric equations for the Hopfield model} J. Math. Phys. \textbf{40}, 3930-3947 (1999).


\bibitem{ton1} I. Perez-Castillo, B. Wemmenhove, J.P.L. Hatchett, A.C.C. Coolen, N.S. Skantzos and T. Nikoletopoulos, {\em Analytic solution of attractor neural networks on scale-free graphs}, J. Phys. A \textbf{37}, 8789, (2004).

\bibitem{data2}  Y. Singh, A.S. Chauhan, {\em Neural networks in data mining}, J. Theor. and Appl. Inform. Techn. \textbf{5}, 1, (2009).

\bibitem{sollich} P. Sollich, A. Barra, {\em Notes on the polynomial identities in random overlap structures}, J.
Stat. Phys. \textbf{147}, 351, (2012).

\bibitem{talabook} M. Talagrand, \emph{Spin glasses: a challenge for mathematicians. Cavity and
mean field models}, Springer-Verlag, (2003).

\bibitem{talahopfield1}  M. Talagrand, {\em Rigourous results for the Hopfield model with many patterns},
Probab. Th. Relat. Fields \textbf{110}, 177-276 (1998).

\bibitem{talahopfield2} M. Talagrand, {\em Exponential inequalities and convergence of moments in
the replica-symmetric regime of the Hopfield model}, Ann. Probab.
\textbf{38}, 1393-1469 (2000).

\bibitem{ton3} B. Wemmenhove, A.C.C. Coolen, {\em Finite connectivity attractor neural networks}, J.
Phys. A: Math. and Gen. \textbf{36}, 9617, (2003).



\end{thebibliography}
\end{document}